\documentclass[reprint,superscriptaddress,
groupedaddress,
amsmath,amssymb,
aps,
pra,
]{revtex4-2}

\usepackage[pdftex]{graphicx}
\usepackage{dcolumn}
\usepackage{bm}
\usepackage[pdftex]{hyperref}
\usepackage{color}
\usepackage{here}

\begin{document}
  \title{Quantum reservoir probing of quantum phase transitions}
  \author{Kaito Kobayashi}
  \author{Yukitoshi Motome}
  \affiliation{Department of Applied Physics, the University of Tokyo, Tokyo 113-8656, Japan}
  \date{\today}
  \begin{abstract}
    Quantum phase transitions are highly remarkable phenomena manifesting in quantum many-body systems. 
    However, their precise identifications in equilibrium systems pose significant theoretical and experimental challenges. 
    Thus far, dynamical detection protocols employing global quantum quenches have been proposed, wherein transitions are discerned via global nonequilibrium excitations. 
    In this work, we demonstrate that quantum phase transitions can be detected through localized out-of-equilibrium excitations induced by local quantum quenches. 
    While the resulting dynamics after the quench is influenced by both the local quench operation and the intrinsic dynamics of the quantum system, the effects of the former are exclusively extracted using the cutting-edge framework called quantum reservoir probing (QRP). 
    Through the QRP, we find that the impacts of the local quenches vary across different quantum phases and are significantly suppressed by quantum fluctuations amplified near quantum critical points; consequently, phase boundaries are precisely delineated. 
    We demonstrate that the QRP can detect quantum phase transitions in the paradigmatic integrable and nonintegrable quantum spin systems, and even topological quantum phase transitions, all within the identical framework employing local quantum quenches and single-site observables. 
  \end{abstract}

\maketitle

\section*{INTRODUCTION}

Quantum phase transitions emerge as quintessential manifestations of the interplay among quantum many-body phenomena~\cite{Sachdev:Cambridge:2011,Zinn:Oxford:2021}. 
They are characterized by the universal description across microscopically different systems, which has been central to our understanding of quantum matters. 
Nevertheless, precisely detecting these transitions in equilibrium systems poses significant challenges both theoretically and experimentally. 
This has led recent investigations to focus on nonequilibrium dynamics, which is inherently related to quantum phase transitions through dynamical quantum critical phenomena. 
Indeed, nonequilibrium signatures of quantum phase transitions have been identified in a diverse range of systems, spanning integrable, nonintegrable, and topological systems~\cite{Bhattacharyya:SciRep:2015,Roy:PRB:2017,Heyl:PRL:2018,Daifmmode:PRL:2019,Wei:PRB:2019,Titum:PRL:2019,Daifmmode:PRB:2020,Nie:PRL:2020,Haldar:PRX:2021,Bin:PRB:2023,Jacob:SciPost:2023,Daifmmode:PRB:2023,Daifmmode:PRBL:2023,Lakkaraju:arXiv:2023,Lakkaraju:PRA:2024}. 
A prevalent strategy in these explorations involves a global quantum quench, wherein the entire system is initially configured in a designated state and subsequently undergoes time evolution under a specific Hamiltonian. 
Recent experimental advancements in isolating and manipulating quantum systems have facilitated such sophisticated control in various platforms, including trapped ions~\cite{Leibfried:RevModPhys:2003,Blatt:NatPhys:2012,Richerme:Nature:2014,Brian:SciAdv:2017}, ultracold atoms~\cite{Kinoshita:Nature:2006,Bloch:RevModPhys:2008,Bloch:NatPhys:2012,Michael:Science:2015}, Rydberg atoms~\cite{Zeiher:NatPhys:2016,Zeiher:PRX:2017,Bernien:Nature:2017,Guardado:PRX:2018}, and nitrogen-vacancy centers~\cite{Choi:Nature:2017}. 

A pertinent question then emerges regarding the feasibility of dynamically detecting equilibrium quantum phase transitions through local quantum quenches. 
Given that the local quench operation, in contrast to its global counterpart, selectively influences nontrivial excitations or quasiparticles in specific regions characteristic of each quantum phase~\cite{Villa:PRA:2020,Villa:PRAL:2021,Villa:PRA:2021,Schneider:PRR:2021}, the resulting dynamics may potentially unveil signatures of quantum phase transitions. 
However, the complexity lies in isolating and analyzing the response dynamics directly ascribed to the local quench operation. 
Since the dynamics is influenced by both the local quantum quench and the intrinsic dynamics of the quantum system itself, the latter obscures the former, thereby diminishing the potential benefit of local quenches in facilitating access to localized physics. 

In this paper, we reveal that signatures of quantum phase transitions manifest themselves in out-of-equilibrium local observables subsequent to local quantum quenches. 
To this end, we develop the methodology of ``quantum reservoir probing" (QRP)~\cite{Kobayashi:arXiv:2023}, which selectively extracts the effects of the local quantum quench from the dynamics influenced by multiple factors. 
Through the lens of the QRP, we observe that the local quench exerts varied influences on a single-site operator, depending on the inherent characteristics of quantum phases. 
Moreover, at the quantum critical points, these influences are notably suppressed due to the enhanced quantum fluctuations. 
Capitalizing on these phenomena, the QRP demonstrates the robust capability to detect quantum phase transitions, as exemplified in both integrable and nonintegrable models herein. 
Furthermore, notwithstanding the use of local quenches and single-site operators, we demonstrate that the QRP can identify topological quantum phase transitions, which are typically distinguished by the absence of local order parameters beyond the Landau-Ginzburg-Wilson paradigm.
The broad applicability and design flexibility of our QRP establish it as a versatile tool for probing a variety of quantum phenomena, opening the door to further understanding of quantum many-body physics.

\section*{RESULTS}
\noindent
\textbf{Framework of the QRP}

\begin{figure*}[bthp]
  \centering
  \includegraphics[width=\hsize]{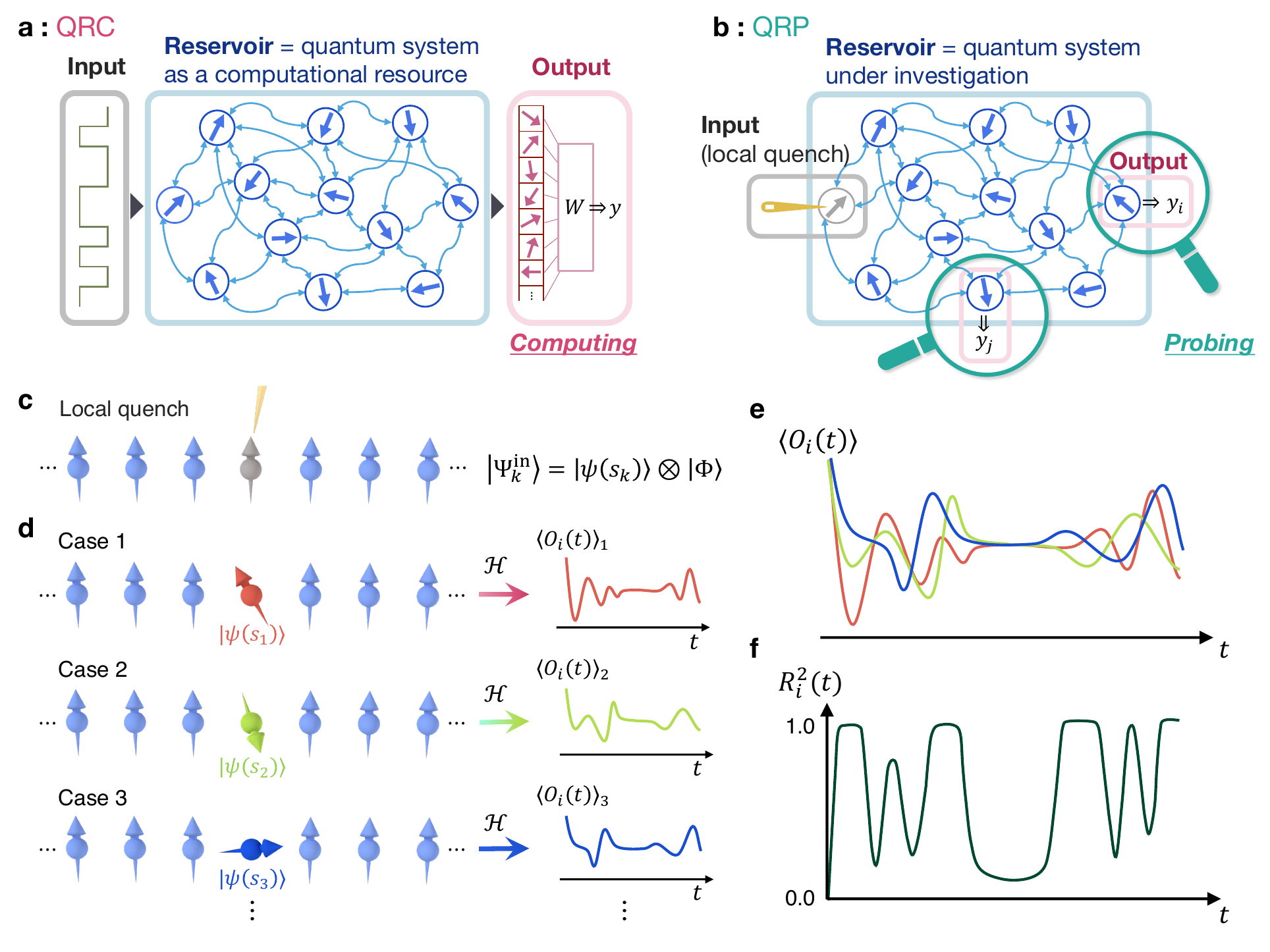}
  \caption{
    \textbf{Conceptualization of the QRP.} 
    \textbf{a} Schematic of the QRC, which harnesses the quantum system as a computational resource. 
    The quantum state of the reservoir reflects the provided input information, and the final output is obtained by linearly transforming a read-out vector constructed through multiple measurements on various degrees of freedom.
    \textbf{b} Schematic representation of the QRP, which investigates the quantum system from the perspective of information processing capabilities. 
    The input is provided via the local quench, the output is calculated from a local degree of freedom, and the estimation performance of the input signifies the impact of the local quench on that degree of freedom.  
    \textbf{c} Preparation and local manipulation to construct the initial state \(|\Psi^{\mathrm{in}}_k\rangle\). 
    The gray spin undergoes the local quantum quench induced by the yellow needle, while the remaining spins are initialized in the designated state \(|\Phi\rangle\), illustrated here as the all-up state. 
    \textbf{d} Diagrammatic representation of the initial quantum state [Eq.~(\ref{eq1})]. 
    The central spin is locally manipulated contingent upon the input value \(s_k\) for each case (three instances are depicted).
    This local quench induces varied dynamics in the local operator \(O_i (t)\), which is reflective of the corresponding input values. 
    \textbf{e} Dynamics of \(\langle O_i (t)\rangle\) summarized across three cases depicted in (\textbf{d}). 
    \textbf{f} Schematic illustration of the determination coefficient \(R_i^2(t)\) derived from \(\langle O_i (t)\rangle\). 
    An almost unity \(R^2_i(t)\) signifies a precise estimation of the input value, indicating the pronounced influence of the local quench on the observed \(\langle O_i (t)\rangle\) [Eq.~(\ref{eq5})]. 
    Conversely, \(R^2_i(t)\) approaches zero when \(\langle O_i (t)\rangle\) exhibits similar values that do not reflect each input, as exemplified by intersections of \(\langle O_i (t)\rangle\). 
    }
  \label{fig1}
\end{figure*}

Before going to the QRP, we provide an overview of a brain-inspired machine learning framework called reservoir computing~\cite{Jaeger:Science:2004,Tanaka:NeuralNetw:2019,Kobayashi:SciRep:2023}. 
In this paradigm, information is nonlinearly processed by networks of artificial neurons, termed the ``reservoir''. 
A distinguishing feature is that all weight parameters within the reservoir part remain fixed, while only the output weight, which transforms the reservoir state into the final output, is trained utilizing a simple linear regression scheme.
Consequently, optimization costs are markedly reduced compared to alternative schemes where the entire network weights are optimized. 
Importantly, due to this fixed nature, a physical system can embody the reservoir part; the quantum reservoir computing (QRC) harnesses a quantum system as the reservoir (Fig.~\ref{fig1}a)~\cite{Fujii:PRA:2017}. 
The quantum reservoir system evolves under the influence of the input data, and a read-out vector, constructed through repeated measurements on various degrees of freedom, undergoes linear transformation to yield the final output. 

The QRP is an inverse extension of the QRC, wherein the quantum system employed as the reservoir part is investigated by linking physical phenomena with information processing (Fig.~\ref{fig1}b). 
As introduced below, the input is provided via the local quantum quench, and the output is computed from a local observable; the objective of the calculation is to estimate the original input value. 
Given that information is provided solely through this local quench, the estimation succeeds when the local quench exerts an influence on the observed degree of freedom. 
By systematically scanning local observation operators, the QRP selectively probes the impacts of the local quench on the quantum system through the input estimation performance. 

To elaborate, the central feature underlying the QRP is the manipulation of a local degree of freedom, parametrized by a random input value \(s_k\). 
Our focus is on a local operator \(O\) defined at spatial position site \(i\) and time \(t\), denoted as \(O_i(t)\), with particular attention to the variations induced by the input variable \(s_k\). 
In the statistical analysis across various instances with different inputs subscripted by \(k\), the expectation value \(\langle O_i (t)\rangle_k\) may display diverse characteristics that closely reflect the corresponding input values. 
In such cases, the value of \(s_k\) can be precisely deduced from \(\langle O_i (t)\rangle_{k}\) through straightforward calculations. 
Hence, the successful estimation of the original input value \(s_k\) provides evidence that the local manipulation exerts an influence on the operator \(O_i (t)\) at that particular spatiotemporal coordinate. 
In contrast, when the operator \(O_j (t')\) at a different spatiotemporal point demonstrates consistent behavior across all instances, it is predominantly governed by the inherent characteristics of the quantum system independent of the local quench. 
Therefore, the dependency of the operator \(O_i(t)\) on the input value \(s_k\) explores how the local quantum quench influences the operator, while isolating it from other intrinsic phenomena.  
Remarkably, this linkage is highly sensitive to the Hamiltonian governing the quantum system. 
The operator's dependency on the local quench operation thus serves as a witness to the properties of the quantum system, shedding light on the underlying quantum processes at play. 

Here, we explain the specific procedures of the QRP on the one-dimensional (1D) quantum systems. 
The protocol consists of four steps: (i) the preparation of the initial state (Fig.~\ref{fig1}c), (ii) the local quench on the central spin, parameterized by the input value \(s_k\) (Figs.~\ref{fig1}c and \ref{fig1}d), (iii) the observation of the dynamics of a local operator \(\langle O_i (t)\rangle\) under the Hamiltonian \(\mathcal{H}\) (Figs.~\ref{fig1}d and \ref{fig1}e), and (iv) the statistical evaluation of the performance in estimating the input value \(s_k\) from the observed dynamics, quantified by the determination coefficient \(R_i^2(t)\) (Fig.~\ref{fig1}f). 
We provide a comprehensive description of each step in the following. 

First, the system is prepared in a designated initial state in step (i); for simplicity, a product state is assumed here (e.g., the all-up state in Fig.~\ref{fig1}c). 
Subsequently, in step (ii), a specific spin undergoes a local quench manipulation based on the input value \(s_k\). 
The resultant state is expressed as 
\begin{equation}
  |\Psi_k^{\mathrm{in}}\rangle=|\psi(s_k)\rangle \otimes |\Phi\rangle,\label{eq1}
\end{equation}
where \(|\psi(s_k)\rangle\) represents the state of the locally manipulated spin, while \(|\Phi\rangle\) denotes the prepared initial state for the remaining subsystem (Fig.~\ref{fig1}d). 
The input value \(s_k\) is randomly sampled from a uniform distribution, specifically \(s_k\in [0, 1]\). 
For instance, we consider the initial state as 
\begin{equation}
  |\psi(s_k)\rangle=\sqrt{1-s_k}|+\rangle + \sqrt{s_k}|-\rangle,\label{eq2}
\end{equation}
where \(|\pm\rangle=\left(\mid\uparrow\rangle\pm\mid\downarrow\rangle\right)/\sqrt2\). 
This state can be prepared by applying a single-gate pulse or a local magnetic field \(\mathbf{h}\propto(2s_k-1,0,-2\sqrt{s_k(1-s_k)})\). 

We then let the system to evolve under the Hamiltonian \(\mathcal{H}\) in step (iii), resulting in the state \(|\Psi_k(t)\rangle=e^{-i\mathcal{H}t}|\Psi_k^{\mathrm{in}}\rangle\) (see Methods section). 
The expectation value of a local operator \(O_i(t)\) is calculated as \(\langle O_i(t)\rangle_{k}=\langle\Psi_k(t)|O_i|\Psi_k(t)\rangle\); we omit the subscript \(k\) unless necessary. 
By iterating this procedure across various instances with different \(s_k\), we generate multiple time series of the dynamics, as illustrated in Fig.~\ref{fig1}e. 

Finally in step (iv), we assess the influence of the local quantum quench on the local operator. 
As a criterion, we employ the precision in estimating the input value \(s_k\) through the linear transformation of \(\langle O_i(t)\rangle_k\), following the standard methodology used in reservoir computing paradigm~\cite{Jaeger:Science:2004,Tanaka:NeuralNetw:2019,Kobayashi:SciRep:2023,Fujii:PRA:2017,Kobayashi:arXiv:2023}. 
For a given instance \(k\), the output of this linear transformation is derived as 
\begin{equation}
  y_{i,k}(t)=w_{i,O}(t)\langle O_i(t)\rangle_k+w_{i,c}(t),
\end{equation}
where \(w_{i,O}(t)\) and \(w_{i,c}(t)\) represent the \(k\)-independent coefficients. 
These weights are finely tuned for training instances so that the output \(y_{{i,k}}(t)\) approximates the original input value \(s_k\) as closely as possible over all \(k\), based on the least squares method.  
Subsequently in the testing phase, the estimation performance for unknown instances is quantified by the determination coefficient
\begin{equation}
  R_i^2(t) = \frac{\mathrm{cov}^2(\{y_{{i,k}}(t)\},\{s_k\})}{\sigma^2(\{y_{{i,k}}(t)\})\sigma^2(\{s_k\})},\label{eq5}
\end{equation}
where \(\mathrm{cov}\) and \(\sigma^2\) represent covariance and variance, respectively; see Methods section for more details. 
The value of \(R^2_i(t)\) ranges between \(0\) and \(1\), with an approach towards \(1\) indicating that the outputs \(\{y_{{i,k}}(t)\}\) closely reproduce the inputs \(\{s_k\}\) for all testing instances.

\begin{figure*}[htbp]
  \includegraphics[width=\hsize]{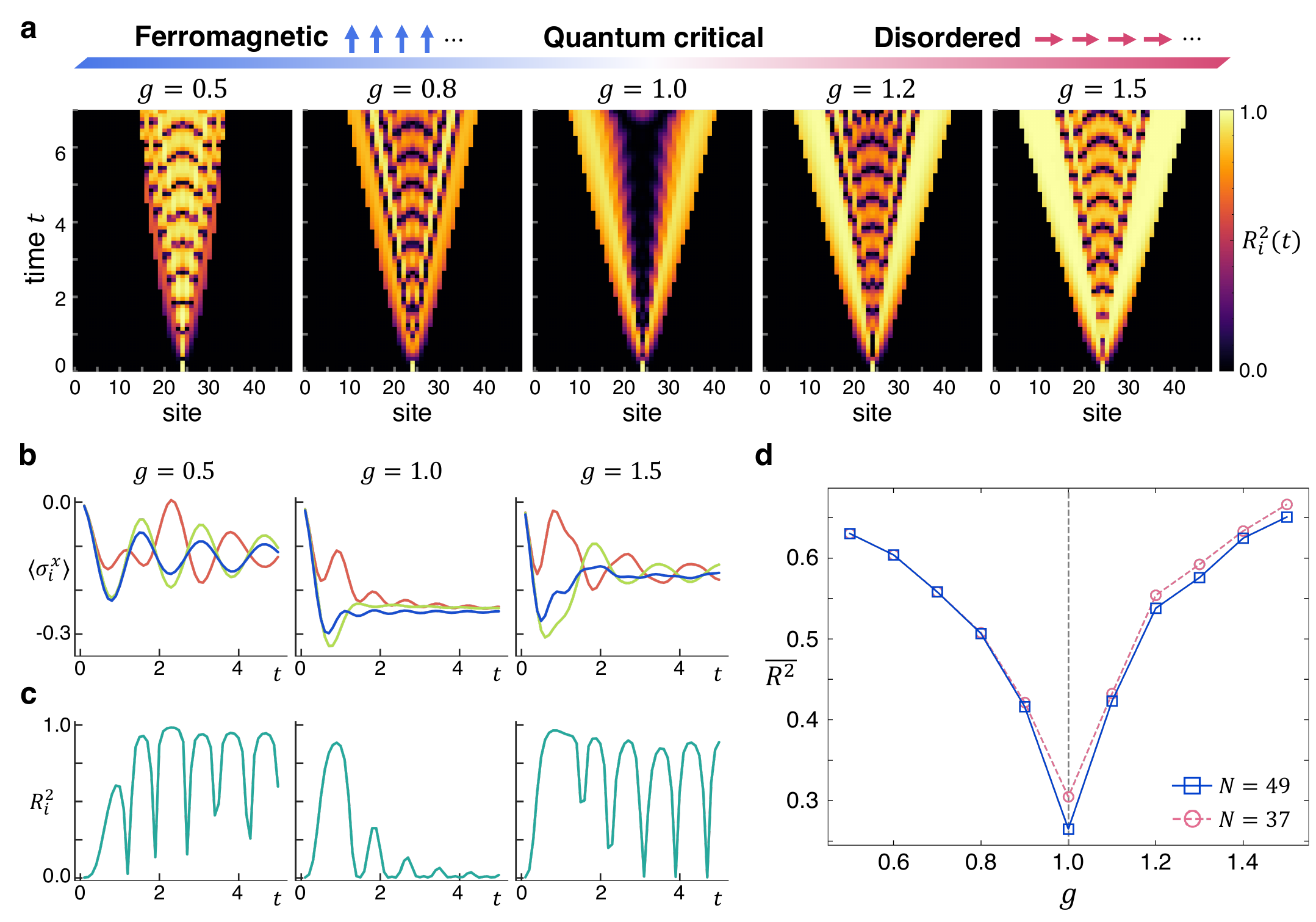}
  \caption{
    \textbf{Identifying the quantum phase transition in the transverse-field Ising model via the QRP.} 
    \textbf{a} Spatiotemporal representation of the performance for the estimation of the input value.
    The color represents the value of \(R^2_i(t)\) derived from \(\langle \sigma^x_i(t)\rangle\). 
    The upper band represents the schematic phase diagram regarding \(\mathcal{H}(g)\), where the quantum critical point separating the ferromagnetically ordered phase and the quantum disordered phase exists at \({g_c}=1\). 
    The system size is \(N=49\), and the time evolution is calculated using a bond dimension \(\chi=256\); the result is quantitatively the same for \(\chi=128\).  
    \textbf{b} The early-time spin dynamics at the nearest-neighbor site from the central one under three different input values. 
    \textbf{c} The estimation performance \(R^2_i(t)\) utilizing \(\langle \sigma^x_i(t)\rangle\) at the same site as (\textbf{b}).  
    \textbf{d} The average \(\overline{R^2}\) over spatiotemporal indices \((i,t)\in\Lambda\times\{0<t\leq 7\}\), with \(\Lambda\) encompassing the central nine sites. 
    The blue and pink lines represent the results for system sizes \(N=49\) and \(N=37\), respectively, exhibiting analogous behavior. 
    The quantum critical point is denoted by the gray dashed line. 
  }
  \label{fig2}
\end{figure*}

Figures \ref{fig1}e and \ref{fig1}f visually elucidate the relationship between the dynamics of \(\langle O_i(t)\rangle_k\) and the corresponding \(R^2_i(t)\). 
A high \(R^2_i(t)\) value suggests that \(\langle O_i(t)\rangle_k\) varies systematically in response to the input value \(s_k\). 
In contrast, \(R^2_i(t)\) close to \(0\) indicates that \(\langle O_i(t)\rangle_k\) is largely independent of \(s_k\), such as at the crossing points resulting from intrinsic oscillations in the quantum system. 
Therefore, \(R^2_i(t)\) serves as an indicator of the extent to which \(O_i(t)\) is influenced by the local quench operation. 
Under different parameters of the Hamiltonian \(\mathcal{H}\), the local quench operation exerts varying influences on the resultant dynamics, hence the behavior of \(R^2_i(t)\) is instrumental in examining the unique properties of each quantum phase. 
It is worth noting that, although the input in Eq.~(\ref{eq2}) includes a nonlinear contribution from \(s_k\), our analysis concentrates on the spatiotemporal dynamics of the linear component in accordance with the QRC framework. 
Any additional complexity beyond linear regression could impede interpretability, as the output is influenced by the extraneous characteristics of the transformation itself. 

\vskip\baselineskip
\noindent
\textbf{Transverse-field Ising model}

Let us first apply our QRP protocol to the paradigmatic 1D Ising model with a transverse magnetic field under the open boundary condition. 
The Hamiltonian is given by
\begin{equation}
  \mathcal{H}=-J\sum_{i}\sigma^z_i\sigma^z_{i+1}+g\sum_{i}\sigma^x_i,
\end{equation}
where \({\sigma}_i^x\) and \({\sigma}_i^z\) are the \(x\) and \(z\) Pauli matrices at site \(i\), respectively. 
\(g\) signifies the magnitude of the transverse magnetic field and \(J>0\) denotes the strength of the nearest-neighbor ferromagnetic interaction; we set \(J=1\) as our energy scale. 
This model manifests a quantum phase transition at the quantum critical point \(g_c=1\) in the thermodynamic limit, which separates a ferromagnetically ordered phase for \(g<1\) from a quantum disordered phase for \(g>1\)~\cite{Coldea:Science:2010,Sachdev:Cambridge:2011}. 
Notably, the model is solvable through the Jordan-Wigner transformation and the subsequent Bogoliubov transformation, where the spin system is mapped to the free quasiparticle picture~\cite{Sachdev:Cambridge:2011}. 

In step (i), the system is initialized in the all-up state, and after the local quantum quench at the central spin in step (ii), the resultant state \(|\Psi^{\mathrm{in}}_k\rangle\) in Eq.~(\ref{eq1}) is explicitly expressed as
\begin{equation}
  |\Psi^{\mathrm{in}}_k\rangle=\mid \uparrow\uparrow\uparrow\dots\rangle\otimes|\psi(s_k)\rangle\otimes\mid \uparrow\uparrow\uparrow\dots\rangle,\label{eq6}
\end{equation}
using \(|\psi(s_k)\rangle\) defined in Eq.~(\ref{eq2}). 
In step (iii), we monitor the dynamics of the local spin \(\langle \sigma^x_i(t)\rangle\) at each site \(i\) under the Hamiltonian \(\mathcal{H}=\mathcal{H}(g)\). 
Repeating the procedure for many instances with varying \(s_k\), the estimation performance \(R^2_i(t)\) in Eq.~(\ref{eq5}) is methodically evaluated in step (iv). 

Figure \ref{fig2} encapsulates the results of the QRP under various magnetic fields \(g\). 
In Fig.~\ref{fig2}a, we show the spatiotemporal representation of the estimation performance \(R^2_i(t)\), which reveals a discernible light-cone structure emanating from the quenched central site. 
The broad distribution of nonzero \(R^2_i(t)\) suggests that the local spin operator \(\langle\sigma_i^x(t)\rangle\) across diverse spatial locations and temporal moments adequately reflect the impact of the local quantum quench. 
In the quantum disordered phase, where the ground state exhibits spins aligned along the \(x\) direction, the effects of the local quench along the \(x\) axis in spin space propagate akin to a spin wave, characterized by pronounced wavefronts of \(R^2_i(t)\). 
In contrast, in the ferromagnetically ordered phase, the local quenching effects are preserved predominantly within the quasiparticles, manifesting as alterations in phase and amplitude. 
This influence is evidenced by the ripple-like patterns of the nonzero \(R^2_i(t)\). 
An exception occurs near the quantum critical point at \(g_c=1\), characterized by maximized fluctuations that lead to an effective breakdown of the quasiparticle picture. 
Here, the effects of the local quench are passed on the intrinsic fluctuations of the system, though such influences rapidly diminish as indicated by the noticeably small \(R^2_i(t)\) after the passage of the wavefront. 
This observation highlights an anomalous suppression of the local quenching effects at the quantum critical point, which is distinct from those in other non-critical phases. 

In Fig.~\ref{fig2}b, the early-time dynamics of \(\langle\sigma^x_i(t)\rangle\) at the nearest-neighbor spin from the central one is depicted for three distinct input values \(s_k\) at different magnetic fields \(g\). 
At \(g=0.5\) and \(1.5\), periodic spin oscillations are induced, with their phase and amplitude dependent on the input \(s_k\). 
In contrast, at \(g=1.0\), the dynamics of \(\langle\sigma^x_i(t)\rangle\) exhibits significant variation when the wavefronts pass at approximately \(t \simeq 1\), thereafter stabilizing into a featureless pattern with diminished influence from the quench operation. 
Figure \ref{fig2}c displays the corresponding estimation performance at the same site (a vertical cross-section of the color map in Fig.~\ref{fig2}a). 
Over broad temporal spans at \(g=0.5\) and \(1.5\), \(R^2_i(t)\) becomes close to \(1\), which statistically underscores the input dependency of the phase and amplitude of the spin oscillations. 
In contrast, their frequency is governed not by the local quench on the initial state, but rather by the intrinsic characteristics of the quasiparticles themselves. 
This is evidenced by \(R^2_i(t)\) periodically approaching zero (Fig.~\ref{fig2}c) at points where oscillations of \(\langle\sigma_i^x(t)\rangle\) intersect for different inputs (Fig.~\ref{fig2}b). 
Notably, at \(g=1.0\), while \(R^2_i(t)\) nears \(1\) around \(t \simeq 1\) where the wavefronts induce variations in the spin dynamics, \(R^2_i(t)\simeq 0\) in other temporal regions signify the reduced influence of the local quench due to the intrinsic fluctuations enhanced near the quantum critical point.

As described above, the effects of the local quantum quench on the resultant dynamics of local operators manifest qualitatively different characteristics across various phases (Fig.~\ref{fig2}a). 
The distribution of \(R^2_i(t)\) thus serves as a marker for identifying the quantum phase transition. 
To enhance the clarity of the results, in Fig.~\ref{fig2}d, we display the mean value of \(R^2_i(t)\) over a specific subset of spatiotemporal indices \(\{(i,t)\}\), denoted as \(\overline{R^2}\). 
This analysis focuses on \((i,t)\) within the central nine sites during the time interval \(0<t\leq 7\); selecting different subsets does not qualitatively alter the results (see the Supplementary Information). 
We note that \(\overline{R^2}\) represents a simple average of local quantities, and therefore, does not include nonlocal quantum effects. 
As \(g\) approaches \(g_c=1\) from the ferromagnetically ordered state, the fluctuations progressively intensify and interfere with the estimation of the original input, resulting in a decrease in \(\overline{R^2}\). 
At the quantum critical point, \(\overline{R^2}\) reaches its minimum, which corresponds to the emergence of a dark region between the wavefronts observed in Fig.~\ref{fig2}a for \(g=1.0\). 
Upon further augmentation of \(g\) into the quantum disordered phase, the fluctuations weaken again and the effects of the quench operation become dominant, leading to an increase in \(\overline{R^2}\). 
The pronounced dip in Fig.~\ref{fig2}d is therefore a definitive indicator of the quantum phase transition occurring at \(g=g_c\). 
Consequently, by focusing on the effects of the local quantum quench, our QRP successfully detects the quantum phase transition only by measurements of single-site spin dynamics.

\begin{figure}[tbhp]
  \includegraphics[width=\hsize]{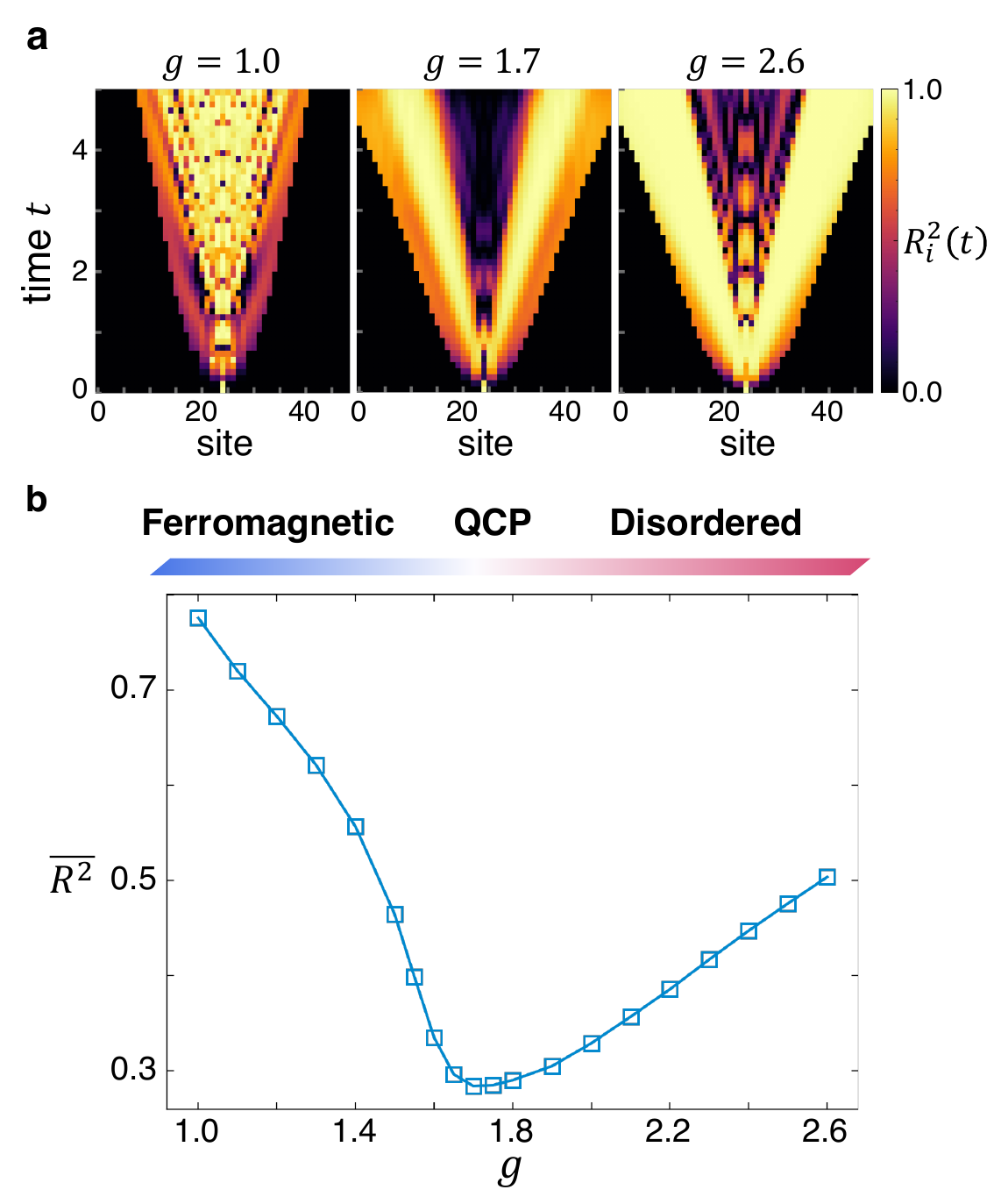}
  \caption{
    \textbf{Detection of the quantum phase transition in the ANNNI model using the QRP.} 
    \textbf{a} Spatiotemporal representation of the estimation performance when employing \(\langle \sigma^x_i(t)\rangle\) in the nonintegrable ANNNI chain characterized by the interaction parameter \(\kappa = 0.5\). 
    The system size is \(N=49\), and the time evolution is calculated with a bond dimension of \(\chi=256\). 
    \textbf{b} The average \(\overline{R^2}\) over a subset \((i,t)\in\Lambda\times\{0<t\leq 5\}\), where \(\Lambda\) contains the central seven sites. 
    The upper band represents the schematic phase diagram, where the ferromagnetically ordered and quantum disordered phases are separated by the quantum critical point at approximately \(g_c\simeq 1.7\).
  }
  \label{fig3}
\end{figure}

\vskip\baselineskip
\noindent
\textbf{Anisotropic next-nearest-neighbor Ising model}

To illustrate the applicability of our QRP framework to nonintegrable systems, we extend the model to the anisotropic next-nearest-neighbor Ising (ANNNI) chain~\cite{Selke:PhysRep:1988,Beccaria:PRB:2006,Chandra:PRE:2007,Suzuki:2013}.
The Hamiltonian is given by
\begin{equation}
  \label{ANNNI}
  \mathcal{H}=-J\sum_{i}\sigma^z_i\sigma^z_{i+1}-\kappa\sum_{i}\sigma^z_i\sigma^z_{i+2}+g\sum_{i}\sigma^x_i,
\end{equation}
where \(J\) and \(\kappa\) represent the strengths of the nerarest-neighbor and the next-nerarest-neighbor interactions, respectively; we take \(J=1\). 
The latter term makes the system nonintegrable by introducing the four-body interactions for the quasiparticles derived by the Jordan-Wigner transformation. 
This model hosts a rich phase diagram in the \(\kappa\)-\(g\) plane~\cite{Karrasch:PRB:2013}. 
Our investigation focuses on the case of \(\kappa=0.5\), where the quantum phase transition occurs at \(g=g_c\) with the Ising universality, separating ferromagnetically ordered (\(g<g_c\)) and quantum disordered (\(g>g_c\)) phases. 
In our finite-size system with open boundary conditions, we determined \(g_c \simeq 1.7\) utilizing the density matrix renormalization group (DMRG) method. 
The initial state is prepared in the same manner as Eq.~(\ref{eq6}). 

Figure \ref{fig3}a displays the spatiotemporal representation of \(R^2_i(t)\) obtained from \(\langle\sigma^x_i(t)\rangle\), revealing a scenario similar to the transverse-field Ising model. 
For the majority of sites and times, the local spin \(\langle\sigma_i^x(t)\rangle\) mirrors the input value \(s_k\) employed in the local quantum quench, as evidenced by the nonzero values of \(R^2_i(t)\).  
Above the critical field at \(g_c\simeq 1.7\), particularly for \(g=2.6\) in the right panel, the effects of the local manipulation propagate primarily through the excitation wavefronts, while the local spin oscillations exhibit weaker dependence on the local quench with small \(R^2_i(t)\). 
Conversely, below the critical point the opposite behavior is noted in the left panel for \(g=1.0\). 
Furthermore, in the vicinity of the quantum critical point, the influence of the local quench manifests itself predominantly along the wavefronts, while in the other regions it is suppressed by the enhanced quantum fluctuations, as demonstrated by the emergence of a discernible dark zone between the wavefronts. 
These fundamental dependencies of the local spin operators on the local quench retain a similarity to those observed in the integrable model presented in Fig.~\ref{fig2}a. 

In Fig.~\ref{fig3}b, we present the mean value \(\overline{R^2}\), calculated over the indices \((i,t)\) corresponding to the central seven sites within the time frame \(0<t\leq 5\) (see the Supplementary Information for other subsets). 
Analogous to the observation in Fig.~\ref{fig2}d, \(\overline{R^2}\) demonstrates a pronounced decrease around the quantum critical point at \(g_c\simeq 1.7\), precisely locating the boundary between the different spatiotemporal distribution patterns of \(R^2_i(t)\) in the ferromagnetically ordered phase and the quantum disordered phase. 
This observation confirms that, even in nonintegrable quantum systems, \(R^2_i(t)\) functions as an effective marker for detecting quantum phase transitions. 

\vskip\baselineskip
\noindent
\textbf{Cluster model}

The preceding sections have demonstrated the effectiveness of our QRP in identifying quantum phase transitions, particularly between conventional phases where the internal physics is well-characterized and understood within the Landau-Ginzburg-Wilson theory. 
Building on this foundation, we now aim to broaden our exploration to more complex quantum phenomena. 
In particular, we apply the QRP framework to the detection of topological quantum phase transitions, which lack local order parameters that can distinguish adjacent phases. 
This raises the question of whether the QRP framework, which solely utilizes a single-site spin operator, can still be effective in identifying these transitions beyond the Landau-Ginzburg-Wilson paradigm.

Here, we study the cluster model with open boundaries~\cite{Raussendorf:PRL:2001,Verresen:PRB:2017,Zeng:2019}, which is defined by
\begin{equation}
  \label{cluster}
  \mathcal{H}=-\sum_{i}J_{zz}\sigma_i^z\sigma_{i+1}^z+\sum_{i}J_{zxz}\sigma_i^z\sigma_{i+1}^x\sigma_{i+2}^z,
\end{equation}
where \(J_{zz}\) and \(J_{zxz}\) denote the Ising and cluster interaction strengths, respectively; we set \(J_{zz}=1\). 
This model is renowned for hosting a symmetry-protected topological (SPT) phase, known as the cluster state, which emerges for \(J_{zz}<J_{zxz}\). 
The SPT phase is distinguished by its protection under a \(\mathbb{Z}_2\times \mathbb{Z}_2^{T}\) symmetry, and is characterized by nonlocal string order parameters~\cite{Perez-Garcia:PRL:2008,Son:EuroPhys:2011,Smacchia:PRA:2011,Pollmann:PRB:2010,Pollmann:PRB:2012}. 
As the strength of the cluster interaction diminishes, the quantum phase transition occurs from the SPT phase to the topologically trivial, ferromagnetically ordered phase at the quantum critical point \(J_{zxz}=1\). 
We aim to detect this transition through the dynamical signature in the QRP, by preparing the initial state in Eq.~(\ref{eq6}) and monitoring the dynamics of the local spin \(\sigma_i^x(t)\). 

\begin{figure}[thbp]
  \includegraphics[width=\hsize]{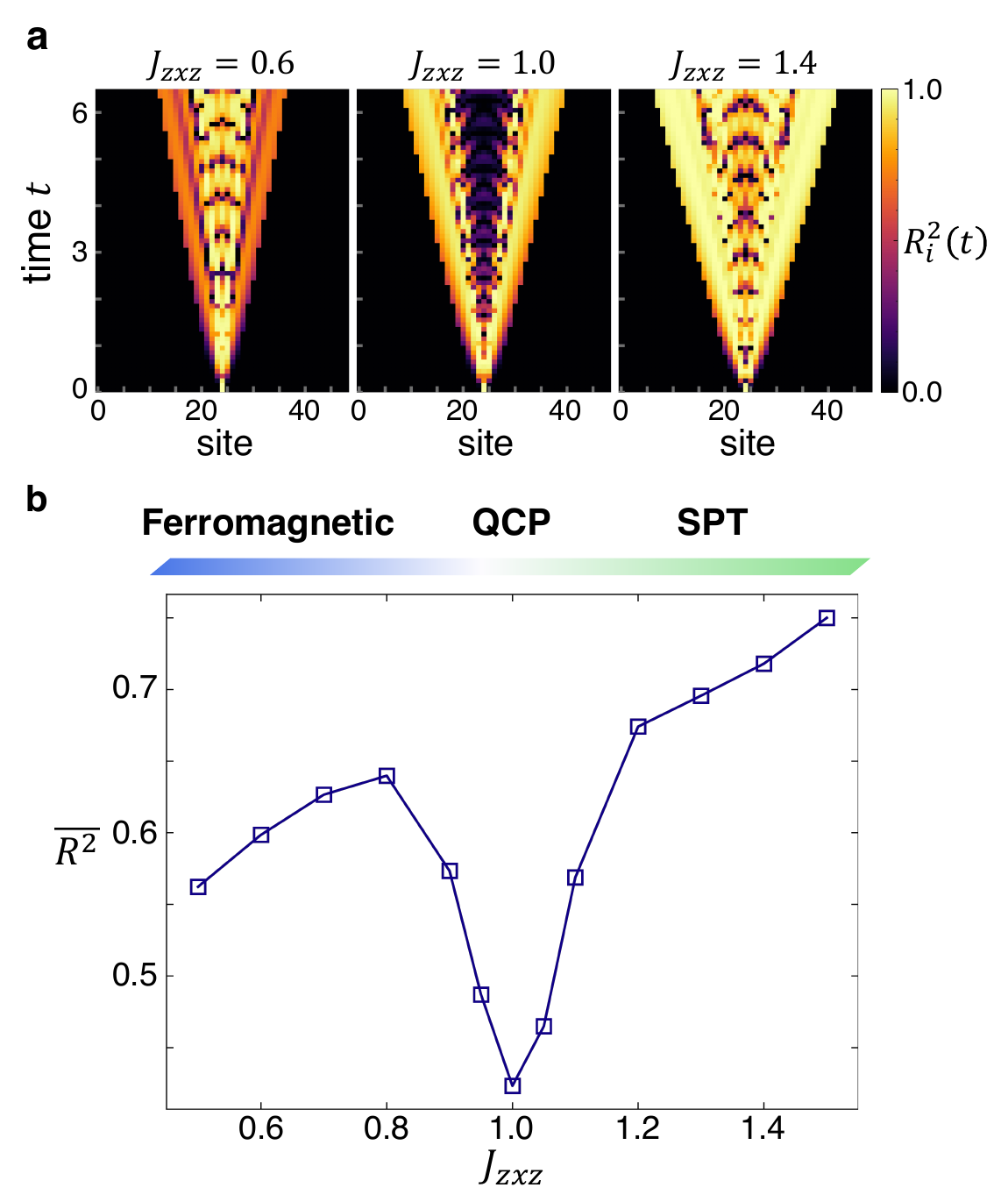}
  \caption{
    \textbf{Probing the quantum phase transition between the ferromagnetically ordered and SPT phases in the cluster model using the QRP.} 
    \textbf{a} Spatiotemporal representation of the estimation performance \(R_i^2(t)\) employing \(\langle \sigma^x_i(t)\rangle\) for a system with \(N=49\) sites. 
    The time evolution is calculated with a bond dimension of \(\chi=256\). 
    \textbf{b} The average \(\overline{R^2}\) over a subset \((i,t)\in\Lambda\times\{0<t\leq6.5\}\), where \(\Lambda\) represents the central eleven spins.  
    The upper bar illustrates the phase diagram.
  }
  \label{fig4}
\end{figure}

In Fig.~\ref{fig4}a, we present the estimation performance \(R^2_i(t)\) under the Hamiltonian \(\mathcal{H}\) for three different strengths of the cluster interaction: \(J_{zxz} = 0.6\) in the ferromagnetically ordered phase, \(J_{zxz} = 1.4\) in the SPT phase, and \(J_{zxz} = 1.0\) at the quantum critical point. 
Consistent with the previous topologically trivial models, the spread of nonzero \(R^2_i(t)\) values in the SPT phase indicates that the operator \( \sigma_i^x(t)\) is reflective of the local quantum quench at various spatiotemporal points. 
Importantly, at the quantum critical point, a notable trend is observed wherein \(R^2_i(t)\) is suppressed following the passage of the wavefronts. 
This precisely mirrors the trend observed at the quantum critical point in conventional quantum phase transitions (Figs.~\ref{fig2}d and \ref{fig3}b).

Figure \ref{fig4}b presents the mean value \(\overline{R^2}\), which effectively distinguishes these three distinct distributions of \(R^2_i(t)\). 
\(\overline{R^2}\) is computed as the average of \(R^2_i(t)\) for the central eleven spins within the time interval \(0<t\leq 6.5\). 
A key observation in Fig.~\ref{fig4}b is the pronounced dip at \(\alpha=\alpha_c\), signifying the suppression of the impact of local quantum quench. 
This dip thus marks the quantum critical point characterized by enhanced quantum fluctuations, clearly delineating the boundary between the SPT phase and the trivial ordered phase. 
Therefore, despite the absence of local order parameters in the topological phase, our QRP, based on the local quench and local observables, demonstrates itself as a potent instrument for detecting the topological quantum phase transition. 

\vskip\baselineskip
\noindent
\textbf{Cluster model in a magnetic field}

Finally, we extend the cluster model in Eq.~(\ref{cluster}) by introducing a transverse magnetic field: 
\begin{equation}
  \label{cluster2}
  \mathcal{H}=\sum_{i}\left[-J_{zz}\sigma_i^z\sigma_{i+1}^z-h_x\sigma_i^x+J_{zxz}\sigma_i^z\sigma_{i+1}^x\sigma_{i+2}^z\right].
\end{equation}
The interplay among these three terms gives rise to a complex phase diagram~\cite{Wolf:PRL:2006,Skrovseth:PRA:2009,Verresen:PRL:2018,Smith:PRR:2022}: the SPT phase is stabilized when the cluster interaction term dominates with a large \(J_{zxz}\), while topologically trivial phases emerge in the ferromagnetically ordered state under strong Ising interaction or in the disordered state under a strong magnetic field. 
Here, we fix \(J_{zz} = 0.1\) and vary \(J_{zxz}=(1-J_{zz})\alpha\) and \(h_{x}=(1-J_{zz})(1-\alpha)\), with a tuning parameter \(\alpha\) adjusting the balance between the cluster and magnetic field terms. 
At the critical value \(\alpha_c=0.5\), the topological quantum phase transition occurs, delineating the SPT phase (\(\alpha > \alpha_c\)) from the disordered phase (\(\alpha < \alpha_c\))~\cite{Smith:PRR:2022}. 
Importantly, these phases are devoid of local orderings on both sides of the quantum critical point, unlike the transition examined in Fig.~\ref{fig4}, which exhibits the local ferromagnetic order on one side.
We initialize the system as
\begin{equation}
  |\Psi^{\mathrm{in}}_k\rangle=|{+}_{y}{+}_{y}\cdots\rangle\otimes|\psi'(s_k)\rangle\otimes|{+}_{y}{+}_{y}\cdots\rangle,
  \label{eq10}
\end{equation}
where \(|\psi'(s_k)\rangle=\sqrt{1-s_k}|{+}_{y}\rangle+\sqrt{s_k}|{-}_{y}\rangle\) and \(|{\pm}_{y}\rangle=\left(\mid\uparrow\rangle \pm i\mid\downarrow\rangle\right)/\sqrt2\). 
\(R^2_i(t)\) is subsequently evaluated from the dynamics of \(\sigma_i^x\). 
These are chosen to clearly demonstrate the QRP by leveraging the flexibility of both the input and output.

\begin{figure}[t!]
  \includegraphics[width=\hsize]{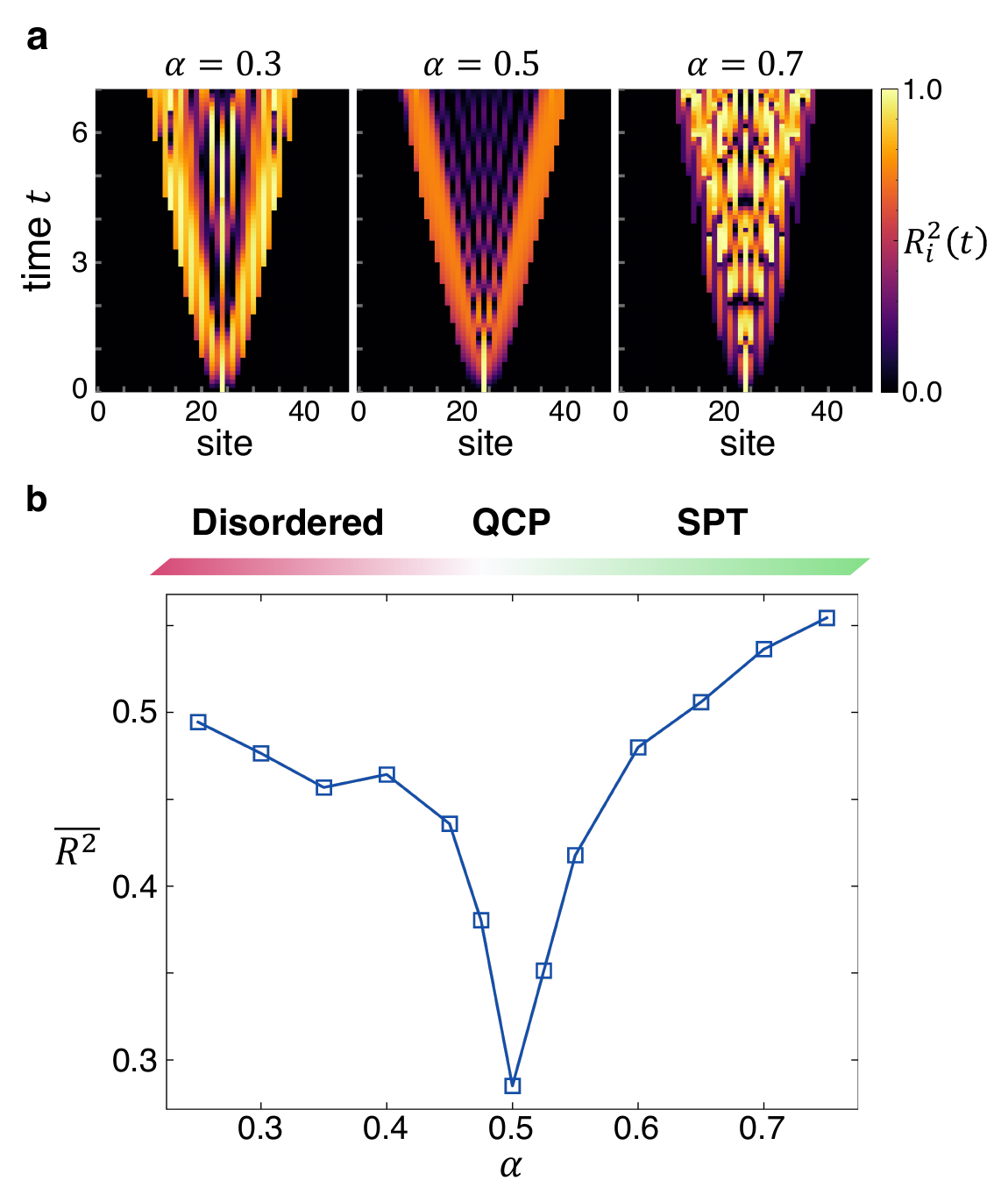}
  \caption{
    \textbf{Detection of the disordered-to-SPT quantum phase transition in the cluster model under a magnetic field via the QRP.} 
    \textbf{a} Color map of the estimation performance \(R_i^2(t)\) for \(\langle \sigma^x_i(t)\rangle\), in the cluster model under a magnetic field with \(J_{zxz}=(1-J_{zz})\alpha\), \(h_{x}=(1-J_{zz})(1-\alpha)\), and \(J_{zz}=0.1\), 
    The system size is \(N=49\), and the time evolution is calculated with a bond dimension of \(\chi=256\). 
    \textbf{b} The average \(\overline{R^2}\) over a subset \((i,t)\in\Lambda\times\{0<t\leq7\}\), where \(\Lambda\) contains the central 13 sites. 
    The upper band represents the phase diagram, where the topological and disordered phases are separated by the quantum critical point at \(\alpha=0.5\). 
    }
  \label{fig5}
\end{figure}

Figure \ref{fig5}a illustrates the estimation performance \(R^2_i(t)\) under the Hamiltonian \(\mathcal{H}\) for three different values of \(\alpha\). 
In all scenarios, \(R^2_i(t)\) acquires nonzero values propagating from the central site throughout the system, where the distinctions between the phases become evident. 
Notably, at the quantum critical point (\(\alpha=\alpha_c\)), \(R^2_i(t)\) is markedly suppressed, displaying a discernibly dark spatiotemporal region in Fig.~\ref{fig5}a. 
This behavior contrasts with the SPT phase at \(\alpha=0.7\) and the trivial disordered phase at \(\alpha=0.3\), both of which exhibits more complicated pattern with widely spread nonzero \(R^2_i(t)\). 
To quantitatively assess these differences, we illustrate the \(\alpha\) dependence of \(\overline{R^2}\) in Fig.~\ref{fig5}b, which is defined by averaging \(R^2_i(t)\) over the central 13 sites within the time frame \(0<t\leq 7\). 
Consistent with the previous models, \(\overline{R^2}\) exhibits a pronounced dip at the quantum critical point, which precisely delineates the boundary between the trivial and topological phases. 
Indeed, the quantum fluctuations are consistently amplified near the critical point, irrespective of the system's topological nature or the existence of local orderings. 
Therefore, even purely topological quantum phase transitions can be detected via the QRP, using these fluctuations in the post-local quench dynamics as witnesses.

\section*{DISCUSSION}

To summarize, we have proposed the QRP as an versatile tool for identifying the quantum phase transitions in a variety of quantum systems, including integrable, nonintegrable, and topological systems. 
The QRP selectively evaluates the effects of the local quench in a manner similar to the pump-probe paradigm. 
Here, the parametrized local quantum quench, acting as the ``pump'', injects random disturbance into the dynamics. 
The estimation performance of this random input parameter, through the linear transformation of the single-site operator, serves as the ``probe''. 
This pump-probe-like process informationally discerns the underlying relationship between the local quench and the resultant dynamics of the operators. 
Through the QRP, we have observed that the local quench operation exerts varied influence depending on the nature of the quantum phases. 
At the quantum critical points, the introduced disturbance becomes subordinate to the maximally enhanced quantum fluctuations. 
A noticeable decline in the estimation performance observed therein serves as a marker for the transition between distinct quantum phases. 
These findings have established that local, nonequilibrium excitations bear the signatures of quantum phase transitions concerning global, equilibrium many-body ground states. 
The crux of our methodology lies in the suppression of local quenching effects on local dynamics due to amplified quantum fluctuations, the rigorous justification of which is left for future study. 

Our investigations have focused on 1D quantum spin systems; however, the QRP methodology is applicable to a broader range of quantum systems in more than one dimension as well. 
Indeed, the conformal field theories governing quantum phase transitions in two- and higher-dimensional systems exhibit significant differences from their 1D counterparts~\cite{Francesco:1997}. 
These differences manifest themselves as changes in nature of excitations, intrinsic dynamics, and critical behaviors near quantum critical points, all of which would be detectable through the QRP. 
Moreover,  geometric frustrations in high-dimensional quantum spin systems, coupled with the inherent quantum fluctuations, give rise to a plethora of topological quantum phase transitions in a variety of systems. 
For instance, in certain systems, such transitions are associated with the emergence of anyons, which have promising potential for applications in topological quantum computing~\cite{Kitaev:2006}. 
The examination of how anyons respond to local quantum operations through the QRP holds significant importance in light of the aforementioned applications. 
We leave the high-dimensional applications of the QRP for future investigations.

Finally, let us remark on the design flexibility of the QRP. 
A fundamental advantage of employing local quantum quenches, in contrast to global ones, lies in their capability to access and elucidate quantum phenomena induced by local operations, such as specific types of excitations. 
When one possesses prior knowledge regarding the physics of the targeted system, especially for discerning partially understood phenomena, the optimal selections for the initial state, local operation, and read-out operators in the QRP become intuitively evident. 
In contrast, the true value of the QRP framework becomes markedly pronounced in the exploration of completely unknown quantum systems. 
Therein, the underlying physics remains undisclosed, rendering the interpretation of the complex dynamics observed a formidable task. 
Our QRP provides a systematic approach to this complexity, categorizing operators by meticulously investigating their dependence on various local operations through a scan of the aforementioned conditions of the QRP. 
Furthermore, the experimental simplicity of the QRP is noteworthy, as it requires solely single-site measurements during the early stages of the dynamics, without the need for complex quantum correlations or long-time evolutions. 
Consequently, we believe that the QRP presents a promising methodology for eliciting novel insights into a broad spectrum of exotic quantum many-body phenomena, extending beyond quantum phase transitions.

\section*{METHODS}

\noindent
\textbf{State preparation and time evolution}

To construct the initial state, we define \(\mathcal{H}_{\mathrm{DMRG}}\) that yields \(|\Psi_k^{\mathrm{in}}\rangle\) as its ground state obtained from the DMRG method~\cite{White:PRL:1992,White:PRB:1993,Schollw:RevModPhys:2005}. 
The Hamiltonian \(\mathcal{H}_{\mathrm{DMRG}}\) comprises the local magnetic fields, aligned with the product state specified in Eqs.~(\ref{eq6}) and (\ref{eq10}), as well as the input magnetic field applied to the central spin to prepare it in the \(s_k\)-dependent state. 
The initial state is then obtained as the ground state of \(\mathcal{H}_{\mathrm{DMRG}}\) using the DMRG method, expressed as a matrix product state with a bond dimension \(\chi\). 
For the time-evolution \(e^{-i\mathcal{H}t}|\Psi_k^{\mathrm{in}}\rangle\), we employ the time-dependent variational principle~\cite{Haegeman:PRL:2011,Haegeman:PRB:2016,Yang:PRB:2020} with a bond dimension \(\chi\) and a time step \(\Delta t = 0.005\). 
To maintain computational accuracy, the time evolution is conducted only up to the point before any anomalous behavior in the entanglement entropy is observed (refer to Supplementary Information). 
The tensor network calculations are implemented using the Julia version of the ITensor library~\cite{Matthew:SciPost:2022,Matthew:SciPostCode:2022}. 

\vskip\baselineskip
\noindent
\textbf{Statistical treatment in the QRP}

From a uniform distribution \([0, 1]\), a set of \(256\) input values \(\{s_k\}\) is sampled. 
Each input \(s_k\) produces a distinct initial state \(|\Psi_k^{\mathrm{in}}\rangle\), and the dynamics of operator \(\langle O_i(t)\rangle_k\) is subsequently calculated. 
Among them, \(l^{\mathrm{tr}}=128\) instances are designated for training, while the remaining \(l^{\mathrm{ts}}=128\) instances are reserved for testing. 
Using a 2D vector \(\mathbf{x}_{i,k}(t) = \left(\langle O_i(t) \rangle_k, 1\right)\), the linear transformation for estimating \(s_k\) is expressed as \(y_{i,k}(t)=\mathbf{x}_{i,k}(t)\mathbf{w}_i(t)\), where the weight vector is defined by \(\mathbf{w}_i(t)=(w_{i,O}(t),w_{i,c}(t))^\top\). 

In the training phase, an (\(l^{\mathrm{tr}}\times 2\))-dimensional matrix \(\mathrm{X}_i^{\mathrm{tr}}(t)=\{\mathbf{x}_{i,k}(t)\}_{k=1}^{l^{\mathrm{tr}}}\) is constructed by summing up the observed dynamics. 
The output, represented as an \(l^{\mathrm{tr}}\)-dimensional vector \(\mathbf{y}_i^{\mathrm{tr}}(t)=\{y_{i,k}\}_{k=1}^{l^{\mathrm{tr}}}\), is calculated as \(\mathbf{y}_i^{\mathrm{tr}}(t)= \mathrm{X}^{\mathrm{tr}}_i(t)\mathbf{w}_i(t)\). 
The weight vector is trained to closely match the output \(\mathbf{y}_i^{\mathrm{tr}}(t)\) with the original input \({\mathbf{{s}}}^{\mathrm{tr}}=\{s_{k}\}_{k=1}^{l^{\mathrm{tr}}}\). 
The optimal solution, which minimizes the least squared error between these vectors, is given by
\begin{equation}
  \mathbf{w}_i(t)={\mathrm{X}_{i}^{\mathrm{tr}}}(t)^+{\mathbf{{s}}}^{\mathrm{tr}},
\end{equation} 
where \({\mathrm{X}_{i}^{\mathrm{tr}}}(t)^+\) denotes the Moore-Penrose pseudoinverse-matrix of \({\mathrm{X}_{i}^{\mathrm{tr}}}(t)\). 

In the testing phase, an (\(l^{\mathrm{ts}}\times 2\))-dimensional matrix \(\mathrm{X}_i^{\mathrm{ts}}(t)=\{\mathbf{x}_{i,k}(t)\}_{k=l^\mathrm{tr}+1}^{l^{\mathrm{tr}}+l^{\mathrm{ts}}}\) is similarly composed, yielding the output \(\mathbf{y}_i^{\mathrm{ts}}(t)= \mathrm{X}_i^{\mathrm{ts}}(t)\mathbf{w}_i(t)\) using the trained weights. 
The estimation performance is assessed by comparing this testing output with the corresponding input \({\mathbf{{s}}}^{\mathrm{ts}}=\{s_{k}\}_{k=l^\mathrm{tr}+1}^{l^{\mathrm{tr}}+l^{\mathrm{ts}}}\) using the determination coefficient 
\begin{equation}
  R^2_i(t) = \frac{\mathrm{cov}^2(\mathbf{y}^{\mathrm{ts}}_i(t),{\mathbf{{s}}}^{\mathrm{ts}})}{\sigma^2(\mathbf{y}^{\mathrm{ts}}_i(t))\sigma^2({\mathbf{{s}}}^{\mathrm{ts}})}. \label{eqM2}
\end{equation}
Here, \(\mathrm{cov}\) and \(\sigma^2\) denote covariance and variance, respectively. 
The coefficient \(R^2_i(t)\) statistically assesses the extent to which \(O_i(t)\) is influenced by the local quantum quench parameterized by \(s_k\). 
We have verified that a smaller dataset with \(l^{\mathrm{tr}}=100\) and \(l^{\mathrm{ts}}=100\) produces quantitatively similar results.

As explained above, the QRP analyzes the variation of dynamics under different \(s_k\) values. 
\(R^2_i(t)\) should ideally be zero when no systematic variations dependent on the input \(s_k\) are observed. 
However, in numerical simulations, \(R^2_i(t)\)  might accidentally exhibit nonzero values, sensitively reflecting the small deviations in \( \langle O_i(t) \rangle_k\) comparable to numerical errors. 
To avoid this numerical artifact, we introduce a threshold \(\Delta_{\mathrm{th}}\) for the deviation of the operator across different instances. 
We employ the average absolute deviation \(\Delta_i(t)\), defined as 
\begin{equation}
  \label{eq13}
  \Delta_i(t) = \frac{1}{l_{\mathrm{tr}}}\sum_{k=1}^{l_{\mathrm{tr}}}\left| \langle O_i(t)\rangle_k - \frac{1}{l_{\mathrm{tr}}}\sum_{k=1}^{l_{\mathrm{tr}}}\langle O_i(t)\rangle_k\right|.
\end{equation}
When \(\Delta_i(t)\) is negligibly small, \(R^2_i(t)\) calculated from the above \(\langle O_i(t)\rangle\) should approach zero. 
We thus manually set \(R^2_i(t)=0\) when \(\Delta_i(t)<\Delta_{\mathrm{th}}\); we utilize \(\Delta_{\mathrm{th}}=10^{-5}\), and \(\Delta_{\mathrm{th}}\) dependency is presented in the Supplementary Information. 
This procedure does not qualitatively alter our methodology and is experimentally feasible.

\section*{DATA AVAILABILITY}
  The data that support the findings of this study are available from the corresponding author upon request.

\section*{CODE AVAILABILITY}
	The simulation codes used in this study are available from the corresponding author upon request.


\section*{ACKNOWLEDGEMENTS}
K.K. thanks Frank Pollmann for insightful discussions. 
This research was supported by a Grant-in-Aid for Scientific Research on Innovative Areas ``Quantum Liquid Crystals" (KAKENHI Grant No.~JP19H05825) from JSPS of Japan and JST CREST (No. JP-MJCR18T2). 
K.K. was supported by the Program for Leading Graduate Schools (MERIT-WINGS), JSPS KAKENHI Grant Number JP24KJ0872, and JST BOOST Grant Number JPMJBS2418. 
The computation in this work has been done using the facilities of the Supercomputer Center, the Institute for Solid State Physics, the University of Tokyo. 

\section*{AUTHOR CONTRIBUTIONS}
K.K. and Y.M. conceived the project. K.K performed the detailed calculations and analyzed the results under the supervision of Y.M. All authors contributed to writing the paper.

\section*{COMPETING INTERESTS}
The authors have no conflicts of interest directly relevant to the content of this article.

\clearpage
\setcounter{equation}{0}
\setcounter{figure}{0}
\setcounter{table}{0}


\renewcommand{\figurename}{Supplementary Fig.}
\renewcommand{\thefigure}{\arabic{figure}}

\widetext
\begin{center}
{\Large Supplementary Information for}

\textbf{\Large Quantum reservoir probing of quantum phase transitions}
\vskip\baselineskip
Kaito Kobayashi\(^*\) and Yukitoshi Motome
\par
{\it Department of Applied Physics, the University of Tokyo, Tokyo 113-8656, Japan} 
\par
(Dated: \today)
\par
\(^*\)Corresponding author. E-mail: kaito-kobayashi92@g.ecc.u-tokyo.ac.jp
\end{center}
\setcounter{page}{1}

\vskip\baselineskip
\begin{center}
\textbf{\large Supplementary Note 1: Average of the estimation performance over various subsets of spatiotemporal indices}\label{Snote1}
\end{center}
\vskip\baselineskip

In the main text, we have demonstrated the efficacy of the quantum reservoir probing (QRP) in identifying quantum phase transitions through a statistical analysis of the effects of the local quantum quench, as described by the estimation performance \(R^2_i(t)\) in Eq.~(5) in the main text. 
Figures 2d and 3b in the main text illustrate the signatures of quantum phase transitions and quantum critical points utilizing the mean value of \(R^2_i(t)\), which is calculated over selected spatiotemporal indices \((i,t)\) to minimize extraneous effects from the edges of the systems. 
Particularly, in the case of the transverse-field Ising model in Fig.~2d, we have presented the field dependence of the mean estimation performance \(\overline{R^2}\), which is obtained by averaging \(R^2_i(t)\) for indices \((i,t)\in \Lambda\times\{0<t\leq 7\}\), where \(\Lambda\) encompasses the central nine sites. 
Similarly for the ANNNI model in Fig.~3b, we have exhibited \(\overline{R^2}\) defined over \((i,t)\) drawn from the subset \(\Lambda\times\{0<t\leq 5\}\), where \(\Lambda\) also contains central seven sites.   
It is imperative to note that the quantum phases manifest distinguishable characteristics in the distribution of \(R^2_i(t)\) (Figs.~2a and 3a in the main text), and \(\overline{R^2}\) serves as an ancillary tool to further clarify these characteristics. 
In other words, this averaging operation is implemented only to enhance the clarity of the results, and \(\overline{R^2}\), defined as the simple average of local quantities, remains inherently local and does not include nonlocal quantum effects. 
In this Supplementary Note, we present \(\overline{R^2}\) defined for various subsets of spatiotemporal indices \((i,t)\) and demonstrate the robustness of the QRP in detecting quantum phase transitions independent of the employed spatiotemporal subset. 

\begin{figure}[b!]
  \centering
  \includegraphics[width=0.9\hsize]{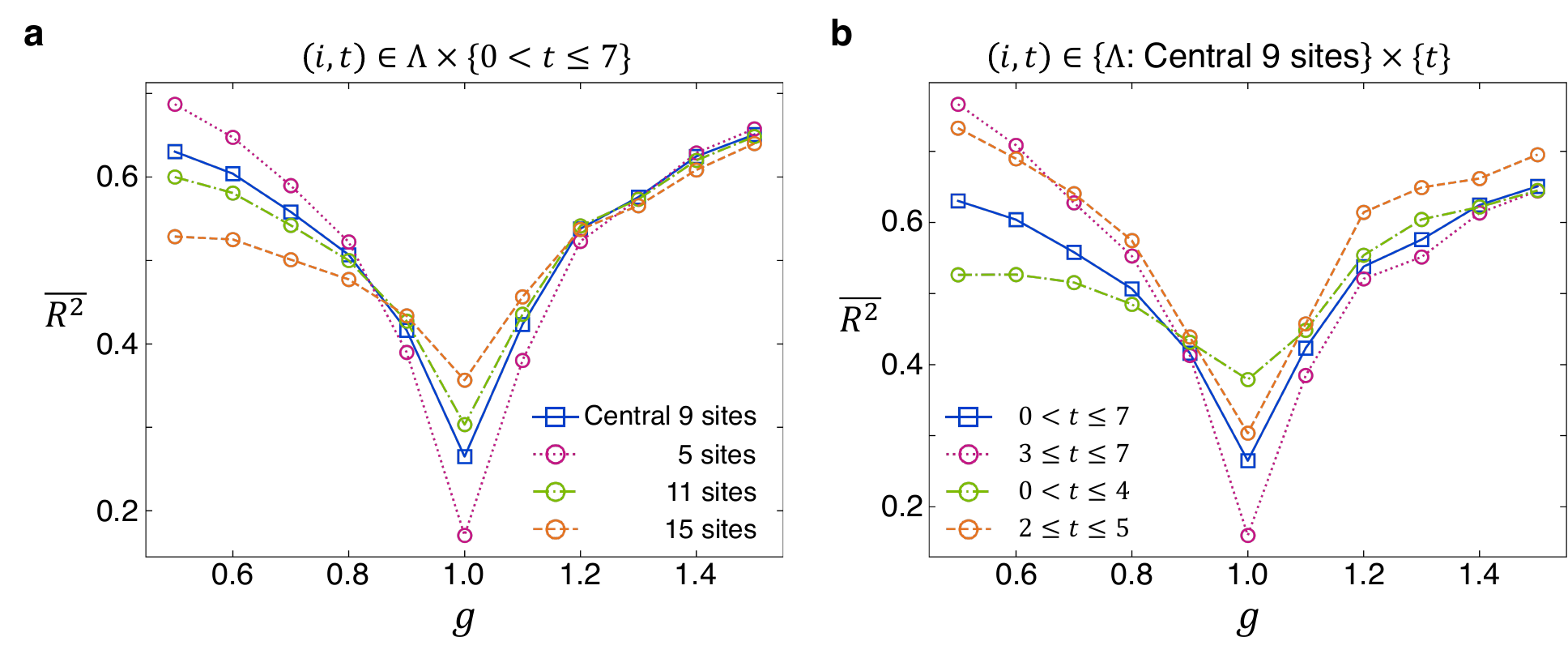}
  \caption{
    \textbf{Average of the estimation performance in the transverse-field Ising model over various subsets of spatiotemporal indices.} 
    \textbf{a} The average \(\overline{R^2}\) over subsets \((i,t)\in\Lambda\times\{0<t\leq 7\}\), where \(\Lambda\) encompasses the central 9 sites (blue), central 5 sites (pink), central 11 sites (green), and central 15 sites (orange). 
    \textbf{b} The average  \(\overline{R^2}\) over subsets \((i,t)\in\Lambda\times\{t\}\), where \(\Lambda\) encompasses the central nine sites. 
    The time window is \(0<t\leq 7\) (blue), \(3\leq t\leq 7\) (pink), \(0<t\leq 4\) (green), and \(2\leq t\leq 5\) (orange). 
    The system size is \(N=49\) in both (\textbf{a}) and (\textbf{b}). 
    }
  \label{Sfig1}
\end{figure}

Supplementary Figure 1 displays the average \(\overline{R^2}\) in the transverse-field Ising model as a function of the field \(g\) in the Hamiltonian \(\mathcal{H}(g)\). 
In Supplementary Fig.~1a, we show \(\overline{R^2}\) computed over various spatial index sets \(\Lambda\), while the temporal index range is held constant at \(\{0<t\leq 7\}\). 
Conversely, in Supplementary Fig.~1b, we present \(\overline{R^2}\) calculated over the fixed set \(\Lambda\) that encompasses the central nine sites, while changing the temporal index range. 
Although each curve corresponding to a different subset \(\{ (i,t)\}\) exhibits a distinct shape, the qualitative characteristics are consistent: \(\overline{R^2}\) decreases as \(g\) approaches \(g_c=1.0\), reaches a minimum at \(g_c\) due to the strong fluctuations, and subsequently rises with further increasing \(g\). 
These patterns robustly indicate the occurrence of the quantum phase transition at the quantum critical point \(g_c\), regardless of the averaging spatiotemporal region. 

Supplementary Figure 2 shows the averaged estimation performance in the ANNNI model as a function of the field strength. 
Supplementary Figure 2a illustrates \(\overline{R^2}\) calculated over various spatial index sets \(\Lambda\) with the temporal index range consistently set at \(\{0<t\leq 5.2\}\). 
In addition, Supplementary Fig.~2b illustrates \(\overline{R^2}\) calculated over a fixed spatial indices set \(\Lambda\) that includes the central five sites, while varying the temporal index range. 
Similar to the observations in the transverse-field Ising model, the qualitative behaviors of \(\overline{R^2}\) in the ANNNI model remain consistent across different subsets \(\{(i,t)\}\), with all results corroborating the occurrence of the quantum phase transition around \(g_c\simeq 1.7\). 
However, a notable feature in the ANNNI model is the broader shape of the dip in \(\overline{R^2}\) near the quantum critical point, which leads to variations in the identified field strength at which \(\overline{R^2}\) reaches its minimum, depending on the specific subset employed. 
Nevertheless, with increasing the system size, it is anticipated that this variance would decrease and the quantum critical point would be identified with greater precision and clarity.

\begin{figure}[t!]
  \centering
  \includegraphics[width=0.9\hsize]{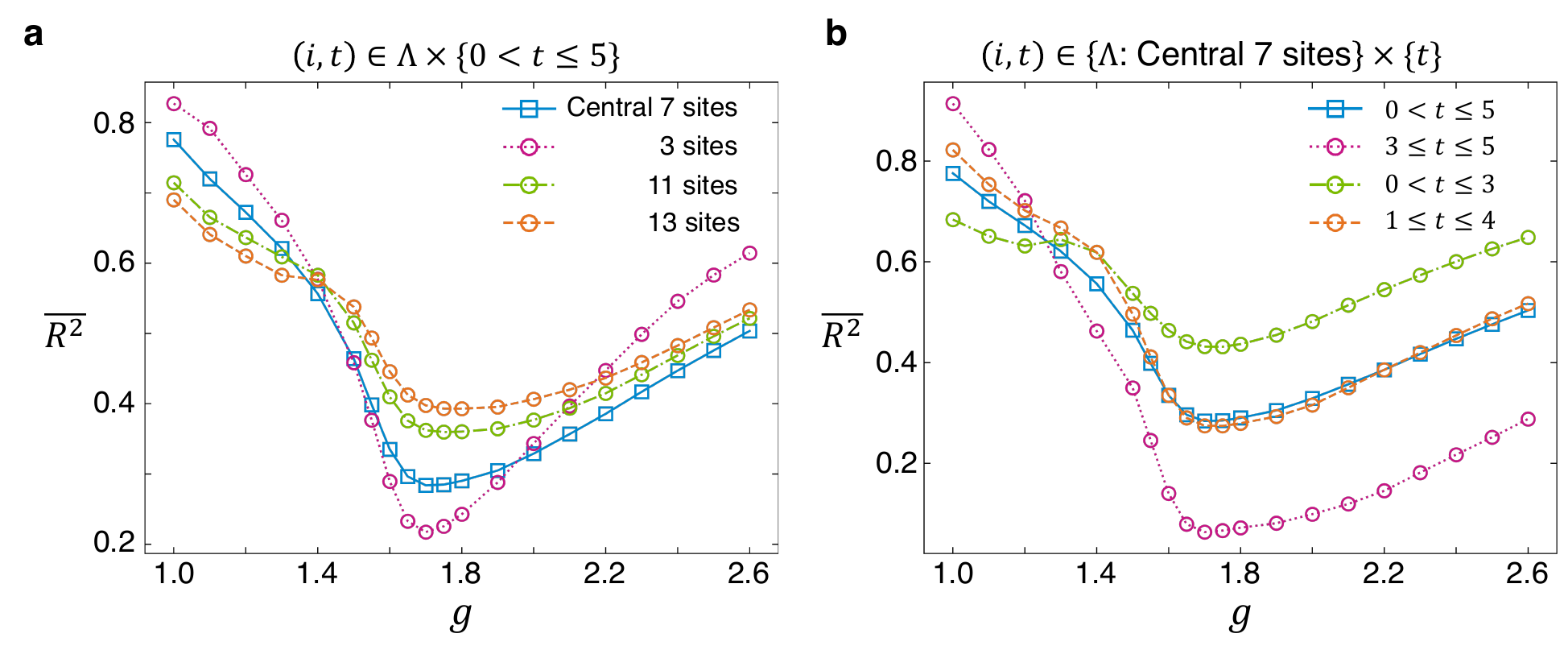}
  \caption{
    \textbf{Average of the estimation performance in the ANNNI model over various subsets of spatiotemporal indices.} 
    \textbf{a} The average \(\overline{R^2}\) over subsets \((i,t)\in\Lambda\times\{0<t\leq 5\}\), where \(\Lambda\) encompasses the central 7 sites (skyblue), central 3 sites (pink), central 11 sites (green), and central 13 sites (orange). 
    \textbf{b} The average  \(\overline{R^2}\) over subsets \((i,t)\in\Lambda\times\{t\}\), where \(\Lambda\) encompasses the central 7 sites. 
    The time window is \(0<t\leq 5\) (skyblue), \(3\leq t\leq 5\) (pink), \(0< t\leq 3\) (green), and \(1\leq t\leq 4\) (orange). 
    The system size is \(N=49\) in both (\textbf{a}) and (\textbf{b}).
    }
  \label{Sfig2}
\end{figure}

\vskip\baselineskip
\vskip\baselineskip
\vskip\baselineskip
\begin{center}
\textbf{\large Supplementary Note 2: Dynamics of the entanglement entropy}\label{Snote2}
\end{center}
\vskip\baselineskip

In the time evolution of a quantum system, the growth of entanglement entropy is a crucial aspect to monitor. 
Specifically, entanglement increases rapidly and spreads throughout the system, eventually exceeding the computational capacity to accurately represent the quantum state beyond a certain temporal threshold. 
This limitation imposes significant constraints on the achievable simulation timescale. 
To explore this, we examine the entanglement entropy between two subsystems, \(A\) and \(B\), defined as \(S=-\mathrm{Tr}_A(\rho_A\ln\rho_A)\), where \(\rho_A\) denotes the reduced density matrix \(\rho_A = \mathrm{Tr}_B\rho\). 
Since a matrix product state with bond dimension \(\chi\) can only capture the entanglement entropy up to \(S\leq\ln\chi\), a temporal threshold arises for precise dynamical calculations, beyond which the state cannot be accurately approximated due to increased entanglement. 
This typically leads to anomalous behavior in the dynamics of the entanglement entropy; for instance, the growth in \(S\) deviates from the expected behavior extrapolated from early times, such as reaching saturation near a value of \(\ln \chi\). 
To ensure numerical accuracy, it is essential to determine the maximum simulation time by closely monitoring this entanglement evolution.

In Supplementary Fig.~3, we present the half-chain entanglement entropy for each model investigated in the main text. 
The initial state \(|\Psi_k^\mathrm{in}\rangle\) in Eq.~(6) or (10) is prepared with a representative input value of \(s_k=0.5\). 
For all models and parameter settings, the entanglement entropy generally shows linear growth during early times, with oscillatory behavior in certain instances. 
At a specific time point (indicated by the gray dashed line in Supplementary Fig.~3), the linear growth rate begins to deviate due to the constraint of maximal entanglement entropy. 
We consider the dynamics beyond this point to be potentially unreliable and focus only on the data prior to the onset of this anomalous behavior. 
Importantly, as demonstrated in the main text, the signatures of the quantum phase transitions become evident as early as \(t\sim1\) or even earlier, which can be captured with high precision both numerically and experimentally.

\begin{figure}[t!]
  \centering
  \includegraphics[width=\hsize]{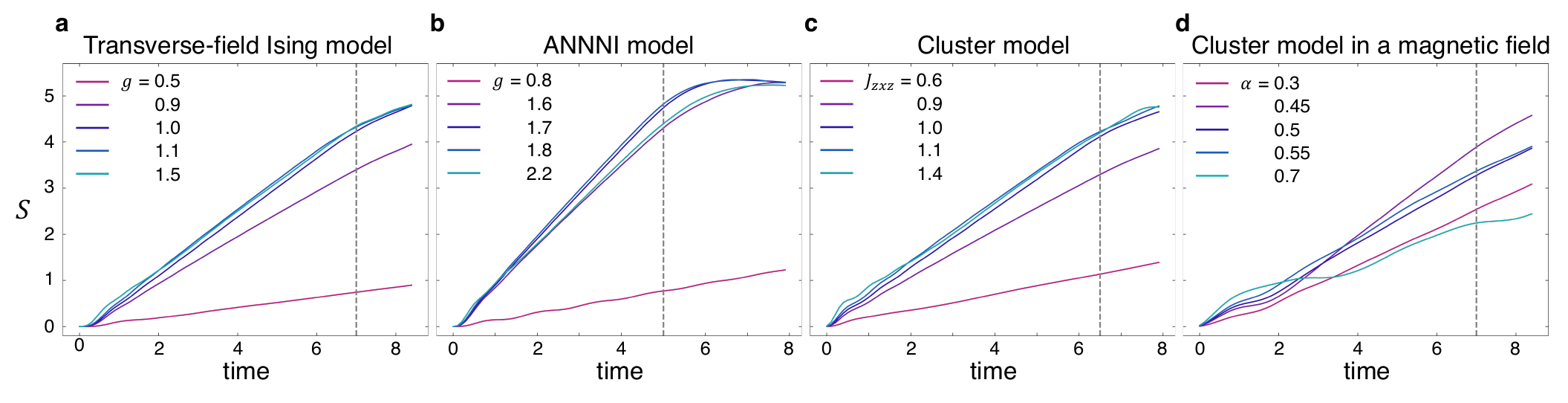}
  \caption{
    \textbf{Dynamics of the entanglement entropy.} 
    The time evolution of the half-chain entanglement entropy \(S\) examined for (\textbf{a}) the transverse-field Ising model, (\textbf{b}) the ANNNI model, (\textbf{c}) the cluster model, and (\textbf{d}) the cluster model in a magnetic field. 
    Each line corresponds to a distinct model parameter. 
    The initial state is prepared with a typical input value \(s_k = 0.5\).  
    The system size is \(N=49\), with a maximum bond dimension of  \(\chi = 256\). 
    The gray dashed line denotes the maximum time utilized to generate the results shown in the main text for each model.
    }
  \label{Sfig3}
\end{figure}

\vskip\baselineskip
\vskip\baselineskip
\begin{center}
\textbf{\large Supplementary Note 3: Dependence of the estimation performance on the deviation threshold}\label{Snote3}
\end{center}
\vskip\baselineskip

As explained in the Methods section in the main text, we introduced the threshold for deviations in the dynamics over multiple instances in the evaluation of the estimation performance. 
The average absolute deviation \(\Delta_i(t)\), defined in Eq.~(13) in the main text, is employed as a criterion. 
\(R^2_i(t)\) should be zero when \(\Delta_i(t)\) is negligible at the level of the numerical error, however, it may accidentally become nonzero, sensitively reflecting the small deviations. 
To circumvent such numerical artifacts, we manually set \(R^2_i(t) = 0\) when \(\Delta_i(t)\) is smaller than the threshold \(\Delta_{\mathrm{th}}\). 
Figures in the main text are obtained under \(\Delta_{\mathrm{th}}=10^{-5}\), whereas the qualitative behaviors remain consistent with different \(\Delta_{\mathrm{th}}\) as shown in the following.

In Supplementary Fig.~4, we present \(R^2_i(t)\) in the transverse-field Ising model for three different \(\Delta_{\mathrm{th}}\); the results with \(\Delta_{\mathrm{th}}=10^{-5}\) are illustrated in Fig.~2a in the main text. 
In Supplementary Fig.~4a, \(R^2_i(t)\) without the threshold is presented, where \(R^2_i(t)\) is depicted as obtained. 
We observe nonzero \(R^2_i(t)\) even outside the light cone, likely due to numerical artifacts.
In Supplementary Fig.~4b and 4c, we introduce the threshold \(\Delta_{\mathrm{th}}=10^{-6}\) and \(\Delta_{\mathrm{th}}=10^{-4}\), respectively. 
The introduction of the threshold narrows the ``width'' of the wavefront in the early stages, and outside the wavefronts the unphysical nonzero values of \(R^2_i(t)\) are eliminated due to reduced numerical sensitivity. 
Once the effects of the local quench have propagated (i.e., inside the wave fronts), the behavior of \(R^2_i(t)\) remains largely unchanged, and the phase-dependent behavior remains consistent.  
Therefore, the introduction of the threshold to the QRP does not fundamentally alter the outcomes of the QRP.

\begin{figure}[t!]
  \centering
  \includegraphics[width=0.85\hsize]{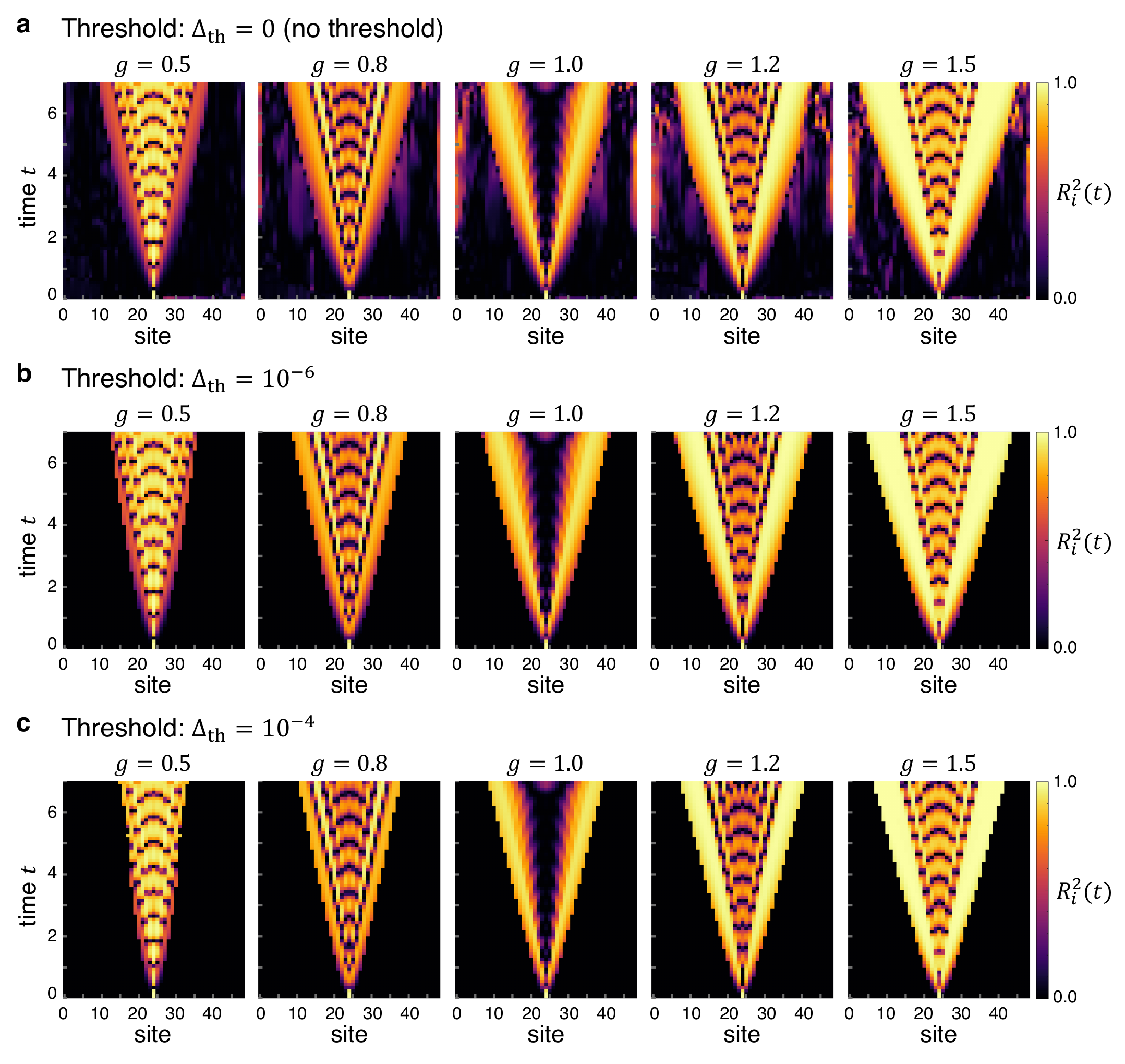}
  \caption{
    \textbf{Dependence of the estimation performance on the deviation threshold in the transverse-field Ising model.} 
    The spatiotemporal representation of \(R^2_i(t)\), as depicted in Fig.~2a in the main text. 
    \(R^2_i(t)\) is calculated with the threshold \textbf{a} \(\Delta_{\mathrm{th}}=0.0\) (no threshold), \textbf{b} \(\Delta_{\mathrm{th}}=10^{-6}\), and \textbf{c} \(\Delta_{\mathrm{th}}=10^{-4}\). 
    }
  \label{Sfig4}
\end{figure}

\end{document}